\documentclass[a4paper,10pt,twoside]{cpc-hepnp}

\usepackage[hang,tight]{subfigure}
\usepackage{threeparttable}
\usepackage{multicol}
\usepackage{graphicx}
\usepackage{booktabs}
\usepackage{float}
\usepackage{amssymb,bm,mathrsfs,bbm,amscd}
\usepackage[tbtags]{amsmath}
\usepackage{lastpage}
\begin{document}

\title{Attenuation length measurements of liquid scintillator with LabVIEW and reliability evaluation of the device\thanks{Supported by the National Natural Science Foundation of China~(11105160, 11005117)}}

\author{%
      GAO Long(¸ßÁú)$^{1,2;1)}$\email{gaol@ihep.ac.cn}%
      \quad YU Bo-xiang(Óá²®Ïé)$^{1;2)}$\email{yubx@ihep.ac.cn(corresponding author)}%
      \quad DING Ya-yun(¶¡ÑÅÔÏ)$^{1}$%
      \quad ZHOU Li(ÖÜÀò)$^{1}$\\%
      \quad WHEN Liang-jian(ÎÂÁ¼½£)$^{1}$%
      \quad XIE Yu-guang(лÓî¹ã)$^{1}$%
      \quad WANG Zhi-gang(ÍõÖ¾¸Õ)$^{1}$%
      \quad CAI Xiao(²ÌÐ¥)$^{1}$\\%
      \quad SUN Xi-lei(ËïÏ£ÀÚ)$^{1}$%
      \quad FANG Jian(·½½¨)$^{1}$%
      \quad XUE Zhen(ѦÕò)$^{1}$%
      \quad ZHANG Ai-wu(Õ°®Îä)$^{1,2}$\\%
      \quad L\"U Qi-wen(ÂÀç²ö©)$^{1,3}$%
      \quad SUN Li-jun(ËïÀö¾ý)$^{1}$%
      \quad GE Yong-shuai(¸ðÓÀ˧)$^{1,2}$%
      \quad LIU Ying-biao(ÁõÓ±±ë)$^{1,2}$\\%
      \quad NIU Shun-li(ţ˳Àû)$^{1,2}$%
      \quad HU Tao(ºúÌÎ)$^{1}$%
      \quad CAO Jun(²Ü¿¡)$^{1}$%
      \quad L\"U Jun-guang(ÂÀ¾ü¹â)$^{1}$%
}
\maketitle

\address{%
1~(State Key Laboratory of Particle Detection and Electronics (Institute of High Energy Physics, CAS), Beijing 100049)\\
2~(Graduate University of Chinese Academy of Sciences, Beijing 100049, China)\\
3~(Shanxi University, Taiyuan 030006, China)\\
}
\begin{abstract}
The attenuation length measuring device was constructed by using oscilloscope and LabVIEW for signal acquisition and processing. The performance of the device has been tested with a variety of ways, the test results show that the set-up has a good stability and high precision (sigma/mean reached 0.4 percent). Besides, the accuracy of the measurement system will decrease by about 17 percent if a filter is used. The attenuation length of gadolinium-loaded liquid scintillator (Gd-LS) was measured as 15.10$\pm$0.35 m where Gd-LS was heavily used in Daya Bay Neutrino Experiment. In addition, one method based on the Beer-Lambert law was proposed to investigate the reliability of the measurement device, the R-square reached 0.9995. Moreover, three purification methods for Linear Alkyl Benzene (LAB) production were compared in the experiment.
\end{abstract}



\addvspace{+0cm}
\begin{multicols}{2}

\section{Introduction}
Liquid scintillator (LS) is applied in many neutrino experiments as the target matter. Linear Alkyl Benzene (LAB) is the most important constituent of LS. Both Palo Verde neutrino experiment and CHOOZ neutrino experiment use LAB as the flashing solvent of LS~\cite{lab1}. In Daya Bay Neutrino Experiment, the new recipe and the manufacturing process were adopted to improve the performance of the gadolinium-loaded liquid scintillator (Gd-LS)~\cite{lab2}. It turns out that the performance of the Gd-LS can meet the requirement of Daya Bay Neutrino Experiment for $\theta_{13}$ measurement~\cite{lab3}. In this work, the set-up based on oscilloscope and LabVIEW was built following the idea that makes it easy to construct as shown in Fig. 1. The LabVIEW program can automatically eliminate the background from every waveform. Otherwise, the accuracy improved due to the use of a filter, the spectral width of light will decrease after it passes through the filter.
In the experiment, high-precision oscilloscope, stable high-voltage power supply (HVPS) and impulse generator were used to improve the precision and stability of the system. A method based on the Beer-Lambert law was proposed to study the reliability of the experimental set-up because there are no samples with explicit value of attenuation length that can be used to test the device. After compared with several samples, we took K$_{2}$Cr$_{2}$O$_{7}$ solution as the standard sample with KOH (0.05mol/L) as the solvent.

\section{The experimental set-up}

Besides electronic equipment, the other parts of the experimental facility were placed in a dark room. A light-emitting diode (LED), with peak wavelength of about 430nm, was used as the light source. The LED powered by a pulse generator, the pulse width was set to be 4.6$\mu$s. The light became pointolite after it passed through the collimator, then, it was focused on the filter and the diaphragm. The aperture diaphragm under the filter was used to get a parallel and very fine beam. The diameter of the light coming out from the aperture was about 2 mm, and it didn't change evidently after passing through the sample. The high-voltage power on the photomultiplier tube (PMT:XP2020)~\cite{lab4} was set to be 1500 V. The sample container was one meter-long cylindrical stainless steel pipe whose capacity was one liter.
\begin{center}
 \vspace{0.5cm}
  \includegraphics[width=7cm]{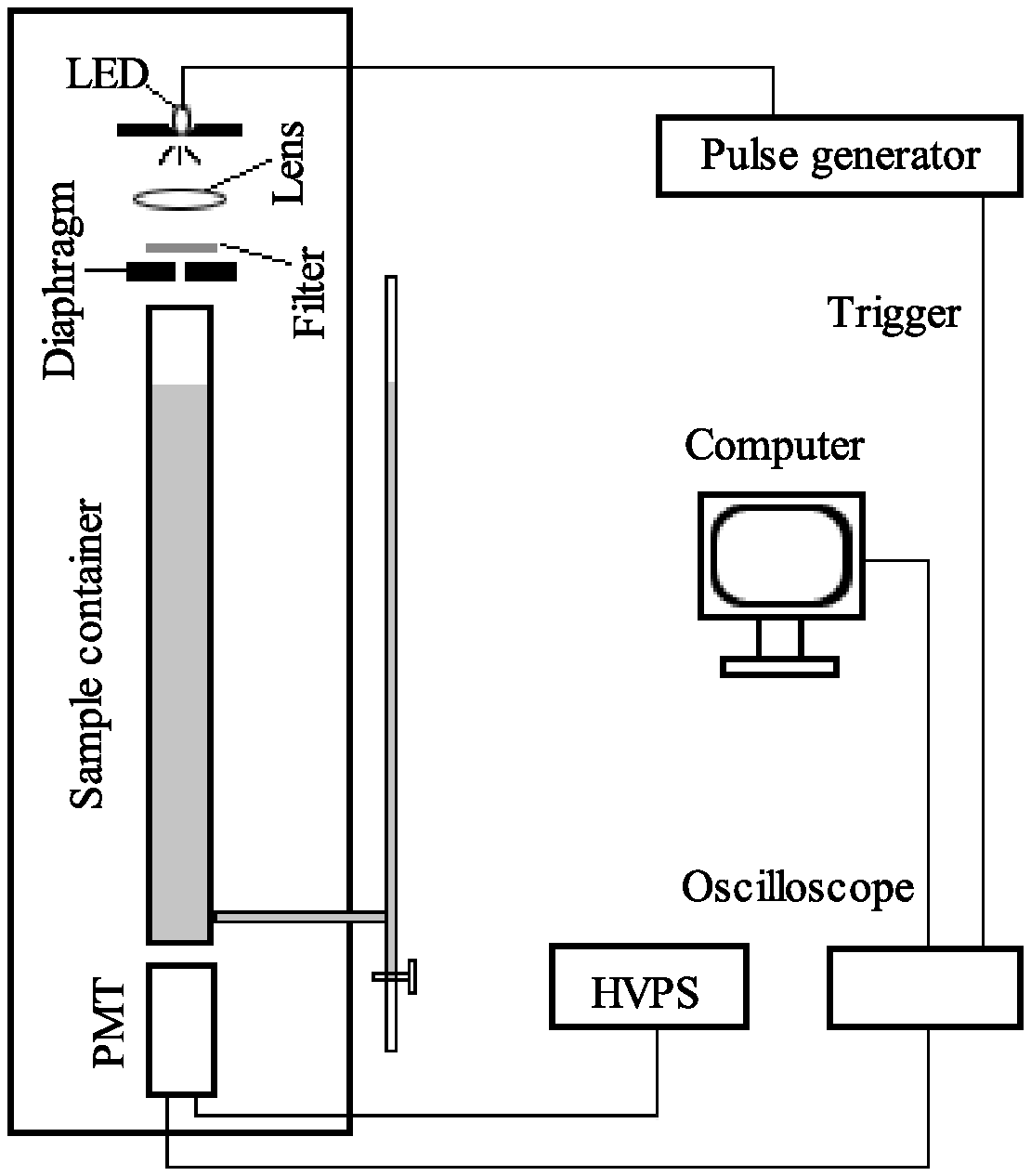}
    \vspace{-0cm}
  \figcaption{\label{fig1} Experimental set up.}
 \end{center}
\section{The experiment}
In physics, the attenuation length or absorption length is defined as the distance ($\lambda$) in a material where the probability of a particle which has not been absorbed has dropped to 1/e. With regard to a beam of particles, the attenuation length is the distance in a material when the proportion of incident particles that have not been absorbed has dropped to 1/e, or about 63 percent of the particles disappeared. The particles are photons produced by LED in this experiment. The relation between I and $I_{0}$ can be expressed as Eq.1.
\begin{equation}
\label{equation}
I=I_{0}{e}^{-\frac{x}{\lambda}},
\end{equation}
where I is the light intensity after it passes through the sample, $I_{0}$ is the original light intensity, x is the path length of light in the sample, and $\lambda$ is the attenuation length. The LabVIEW program can read and process about 300 waveforms from oscilloscope per minute. A waveform consists of ten thousand points, it is transmitted to a computer with reticle and then handled by the LabVIEW program. Fig. 2 is the data flow diagram of the program to acquire and process the signal. The integral of the waveform is the light intensity. The program takes both ends of the waveform as background dynamically. The statistic of each measurement to a light intensity is more than 3000 and the randomness of them fits with Gaussian distribution. For each sample, the intensity of the light changes with the different depths of the sample in the sample container, then the attenuation length can be obtained by fitting the experimental data.
\end{multicols}
\begin{figure}[!htb]
  \begin{center}
  \vspace{-0cm}
          \label{fig2}
          \begin{minipage}[]{1\textwidth}
            \centering
            \includegraphics[width=17cm,height=4cm]{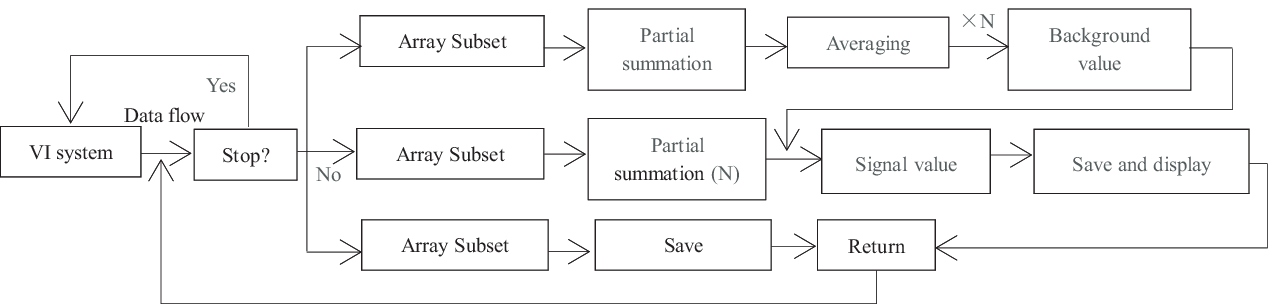}
          \end{minipage}%
        \caption{The data flow diagram of LabVIEW program.}
  \end{center}
\end{figure}
\vspace{-0cm}
\begin{multicols}{2}
\subsection{Stability and precision}

The stability of the device which includes the LED, the pulse generator, PMT, HVPS and the data processing system has been studied. As shown in Fig. 3(a), the change of signal is less than 2.5\% in 48 hours. The time duration of measuring one sample is about 5 hours, so the change is less than 0.3 percent. Therefore, the alterations of the system within an experimental period have little influence on the measurement result. The downswing in Fig. 3(a) can be understood as the PMT will experience tiny and recoverable aging over time. More importantly, as shown in Fig. 3(b), the sigma/mean can reach 0.4 percent after the device is commendably adjusted.
\end{multicols}
\begin{figure}[!htb]
  \begin{center}
  \vspace{-0cm}
        \subfigure[]{
          \label{}
          \begin{minipage}[b]{0.5\textwidth}
            \centering
            \includegraphics[width=7cm]{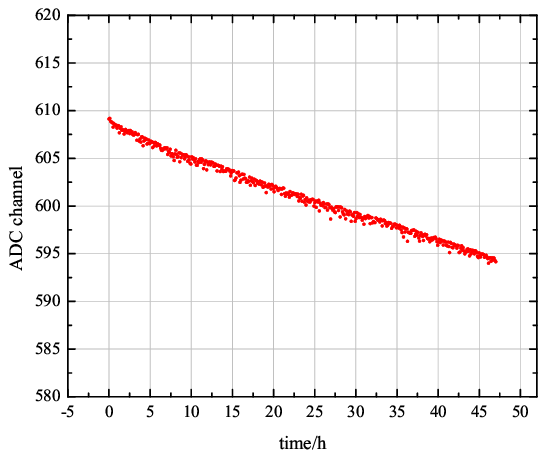}
          \end{minipage}}%
        \subfigure[]{
          \label{}
          \begin{minipage}[b]{0.5\textwidth}
            \centering
            \includegraphics[width=7cm]{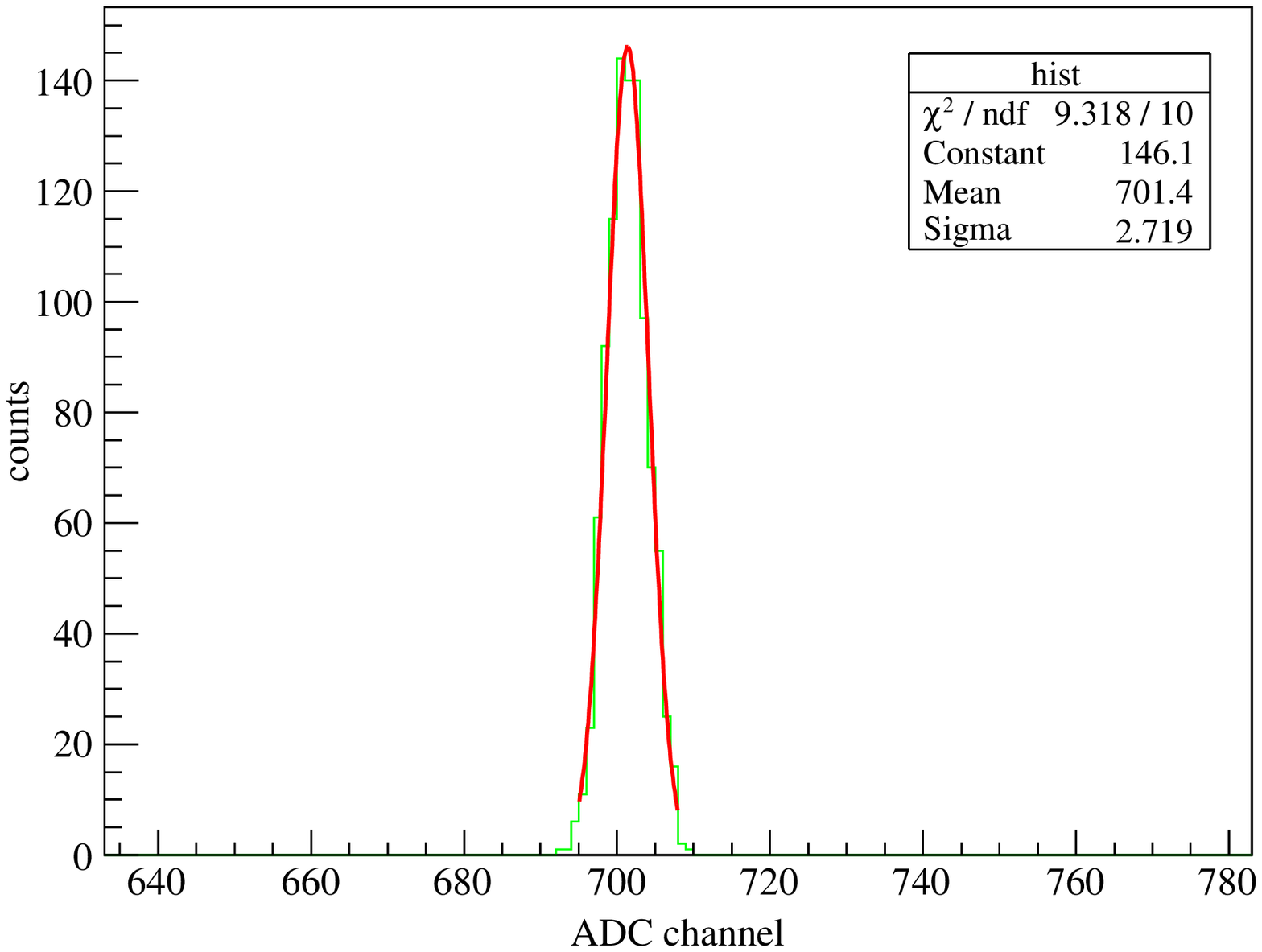}
          \end{minipage}}%
        \caption{The system stability test and data fitting.}
        \label{fig3}
  \end{center}
\end{figure}
\vspace{-0cm}
\begin{multicols}{2}
\subsection{The influence of filter}
The LED is not monochromatic light sources but emit over a significant spectral region. The spectral distribution of the light source used is shown in Fig. 4. The figure also shows that the spectrum width of the beam reduced after it passing through the filter. The following controlled trial will be performed to study how much of an effect the filter will have on the measurements.
\begin{center}
 \vspace{-0cm}
  \includegraphics[width=8cm]{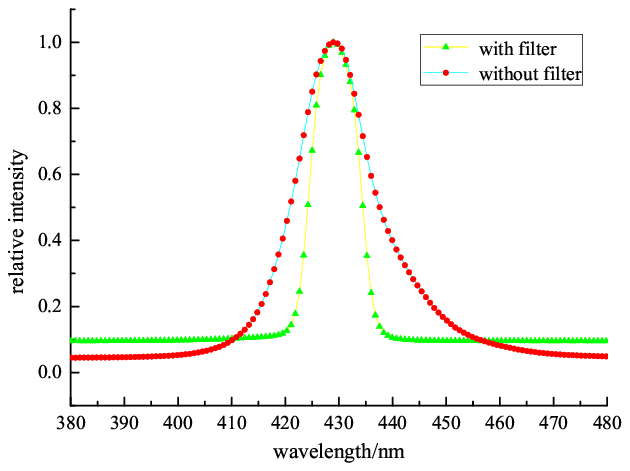}
    \vspace{-0cm}
  \figcaption{\label{fig4} The comparison of LED's spectrum.}
  \vspace{-0cm}
 \end{center}

In the experiment, we tested one kind of LAB for two times. In the first measurement, the attenuation length of the LAB was 13.96$\pm$0.17m, and the value was 13.93$\pm$0.25m in the second time. This indicates the measurement system is stable. When the filter was not used, the attenuation length of the LAB was changed to 17.48$\pm$0.42m. It shows that the spectrum's width of test beam will influence the measurement result and the result will become bigger
\begin{center}
 \vspace{-0cm}
  \includegraphics[width=8cm]{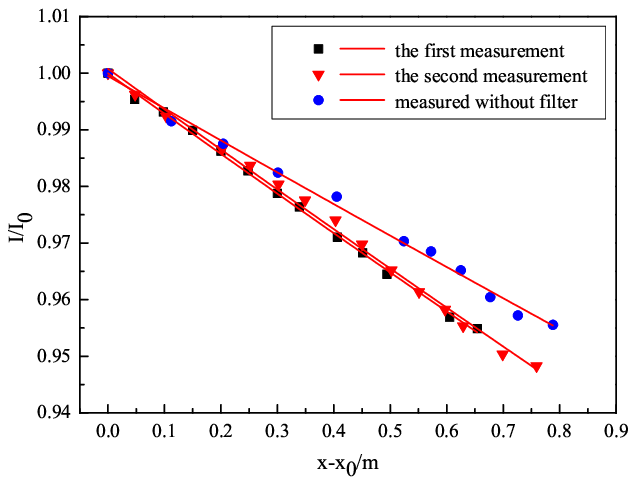}
    \vspace{-0cm}
  \figcaption{\label{fig5} Repeated measurements of LAB and the effect of filter.}
  \vspace{-0cm}
 \end{center}
if the spectrum's width becomes wider. The attenuation length of the sample changes along with the wavelength of the light source according to its definition~\cite{lab5}. So the use of filter makes the results more accurate by narrowing down the spectrum of the light beam. Moreover, the stability of the impulse generator and HVPS is also very important for good results.
\subsection{The reliability evaluation of device}
In this work, a method based on the Beer-Lambert law was proposed to study the reliability of the experimental set-up. The Beer-Lambert law~\cite{lab6} describes the linear relationship between absorbance and concentration of a light-absorbing substance. The general Beer-Lambert law is usually written as:
\begin{equation}\label{equation2}
    A=-\log(I/{I_{0}})={\varepsilon}xc,
\end{equation}
where A is the measured absorbance, $\varepsilon$ is the wavelength-dependent absorptivity coefficient, I is the light intensity after the beam passes through the sample, $I_{0}$ is the initial light intensity, x is the path length of light in the sample, and c is the concentration of analyte. There are several conditions that need to be fulfilled in order to use Beer-Lambert law. For instance, making sure the concentration is lower than 0.01M; ensuring that the incident beams are preferably parallel and monochromatic; balancing in chemical equilibria and stabilization in the concentration. If any of these conditions is not fulfilled, deviations will occur from the Beer-Lambert law~\cite{lab7}.
The uncertainty of the measuring result in the experience can be known by measuring the attenuation length of standard samples and applying the Beer-Lambert law to the measurement results. The standard samples are K$_{2}$Cr$_{2}$O$_{7}$ solution which takes KOH(0.05mol/L) as the solvent. Eq.3 can be obtained by combining the Beer-Lambert law with the definition of attenuation length where 1/$\lambda$ is proportional to the concentration of the standard sample.
\begin{equation}\label{equation3}
    \frac{1}{\lambda}=\varepsilon\ln(10)c.
\end{equation}
There is no low limit in concentration, but at very low concentration, the result of absorbance can be erroneous due to the limited resolution of the measuring system or because the signal-to-noise ratio of the light intensity measurement is too low (due to detector noise, photon noise, or the light source fluctuation). Eq.3 can be applying to the experimental data with linear fit as shown in Fig. 6. The Coefficient of Determination (R-square)~\cite{lab8} can be used as one measure of the quality of the linear regression model, it is 0.9995, which means that the measurement device is reliable~\cite{lab9}.
\begin{center}
 \vspace{-0cm}
  \includegraphics[width=8cm]{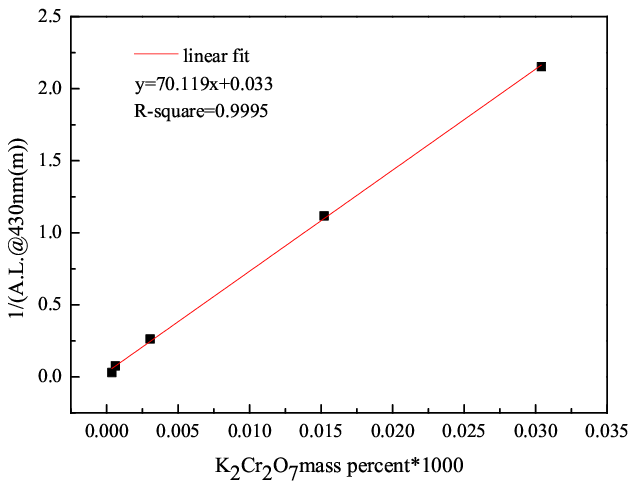}
    \vspace{-0cm}
  \figcaption{\label{fig6} Applying the beer-lambert law to standard samples.}
  \vspace{-0cm}
 \end{center}
\subsection{Attenuation length measurements of the Gd-LS}
Three different samples including LAB, LS and Gd-LS have been measured in this experiment. They are all used in the Daya Bay Neutrino Experiment. Besides LAB, carboxylic acid, gadolinium chloride, PPO and bis-MSB are used in producing Gd-LS. The Gd-LS prepared with such recipe has long attenuation length, high photoyield and long-term stability.
 \begin{center}
 \vspace{-0cm}
  \includegraphics[width=8cm]{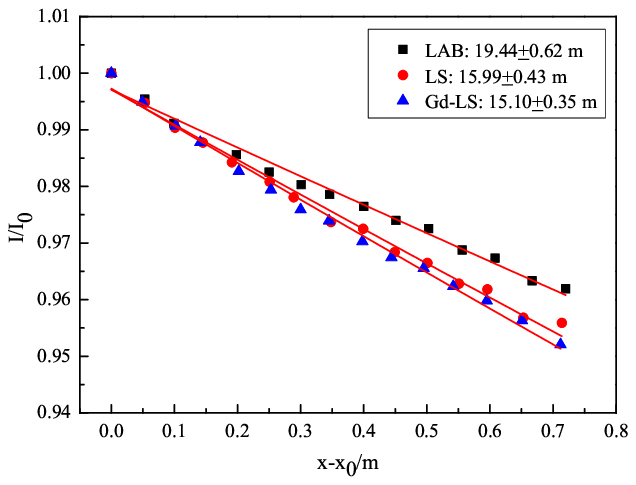}
    \vspace{-0cm}
  \figcaption{\label{fig7} The attenuation length measurements of samples.}
  \vspace{-0cm}
 \end{center}
There was a ruler near the tubule which was connected with the sample container, so the position of the sample would be known. The intensity of the outgoing beam was provided by the LabVIEW program. Choosing a position as the reference, we can get the attenuation length with Eq.1. As shown in Fig. 7, the attenuation length of the LAB is the longest, and the attenuation length of the Gd-LS is 15.10m.
\subsection{Research on purification methods of LAB}
In the Phase II project of the Daya Bay Neutrino Experiment, the detector will become bigger, the PMTs will also be placed near the outer wall of each detector to record the light pulses produced when neutrinos pass through the Gd-LS. If the attenuation length of the Gd-LS is too short, the light cannot come out or the intensity of the light will be too weak to be detected by the PMTs. LAB is the main material to synthesize liquid scintillation. The commercially purchased LAB has been purified by vacuum distillation, aluminum oxide (Al$_{2}$O$_{3}$) column and silica gel column respectively. As shown in Fig. 8, LAB purified by aluminum oxide column has the longest attenuation length.
\begin{center}
 \vspace{-0cm}
  \includegraphics[width=8cm]{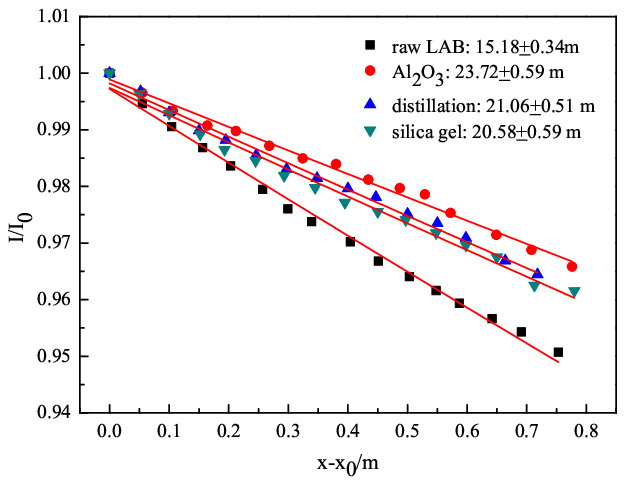}
    \vspace{-0cm}
  \figcaption{\label{fig8} The comparison between different methods in purifying LAB.}
  \vspace{-0cm}
 \end{center}
Table 1 shows the different attenuation lengths of several samples measured in this work. For K$_{2}$Cr$_{2}$O$_{7}$ solution, the unit of concentration is weight percentage.
\begin{center}
\tabcaption{ \label{tab1}  The attenuation lengths of samples.}
\footnotesize
\begin{tabular*}{80mm}{c@{\extracolsep{\fill}}ccc}
\toprule sample & attenuation length ($\lambda$)/m \\
\hline
LAB & \hphantom{0}19.44$\pm$0.62  \\
LS & \hphantom{0}15.99$\pm$0.43  \\
Gd-LS & \hphantom{0}15.10$\pm$0.35  \\
Alcohol & \hphantom{0}16.82$\pm$0.13  \\
0.0365\%K$_{2}$Cr$_{2}$O$_{7}$  & \hphantom{0}32.42$\pm$1.79  \\
0.0611\%K$_{2}$Cr$_{2}$O$_{7}$  & \hphantom{0}13.04$\pm$0.22  \\
0.304\%K$_{2}$Cr$_{2}$O$_{7}$  & \hphantom{0}3.79$\pm$0.04  \\
1.52\%K$_{2}$Cr$_{2}$O$_{7}$  & \hphantom{0}0.89$\pm$0.01  \\
3.04\%K$_{2}$Cr$_{2}$O$_{7}$  & \hphantom{0}0.46$\pm$0.01  \\
\bottomrule
\end{tabular*}
\end{center}
\section{Conclusions}
In this work, the measuring equipment was contrast with LED, filter, PMT, oscilloscope, computer, etc. The LabVIEW program was developed to obtain and handle signal which makes the precision of the whole system reach 0.4 percent. The filter can narrow down the spectrum's width of test beam which enables the test beam to better meet the applicable condition of the formula of attenuation length, therefore the accuracy during measurement will be increased by about 17 percent compared with that without filter. Besides, the reliability of the measure system has been evaluated with a method which is based on the Beer-Lambert law. The result shows that the device is reliable. The Gd-LS and the other several samples have been accurately measured in their attenuation lengths. The attenuation length of Gd-LS is 15.10$\pm$0.35 m, a Monte Carlo parameter of the Daya Bay Neutrino Experiment for $\theta_{13}$ measurement. Moreover, several methods used in purifying LAB have been studied and compared, and it shows that Al$_{2}$O$_{3}$ is the best candidate.

\end{multicols}
\vspace{-1mm}
\centerline{\rule{80mm}{0.1pt}}
\vspace{2mm}
\begin{multicols}{2}

\end{multicols}
\clearpage
\end{document}